%% file: main.tex
\documentclass[sigconf, 10pt]{acmart}

\usepackage{booktabs}
\usepackage{xcolor}
\usepackage{xspace}
\usepackage{tikz}
\usepackage{pifont}
\usepackage{enumitem}

\renewcommand\footnotetextcopyrightpermission[1]{}
\setcopyright{none}
\input{globals}

\acmYear{2027}

\begin{document}
\title{"Is Low Power Enough?" Rethinking Energy Efficiency in High-Fidelity Sensing}

\author{Chae Young Lee}
\orcid{0009-0005-5078-1233}
\affiliation{%
  \institution{Stanford University}
  \city{Stanford} 
  \state{California} 
  \country{USA}
}
\email{chae@stanford.edu}

\author{Lan Mei}
\affiliation{%
  \institution{Stanford University}
  \city{Stanford} 
  \state{California} 
  \country{USA}
}
\email{lanmei@stanford.edu}

\author{Jesse Fonseca}
\affiliation{%
  \institution{Stanford University}
  \city{Stanford} 
  \state{California} 
  \country{USA}
}
\email{fonsecaj@stanford.edu}

\author{Zerina Kapetanovic}
\affiliation{%
  \institution{Stanford University}
  \city{Stanford} 
  \state{California} 
  \country{USA}
}
\email{zerina@stanford.edu}

\renewcommand{\shortauthors}{Lee et al.}

\input{0-Abstract}


\begin{teaserfigure}
  \centering
  \includegraphics[width=\textwidth]{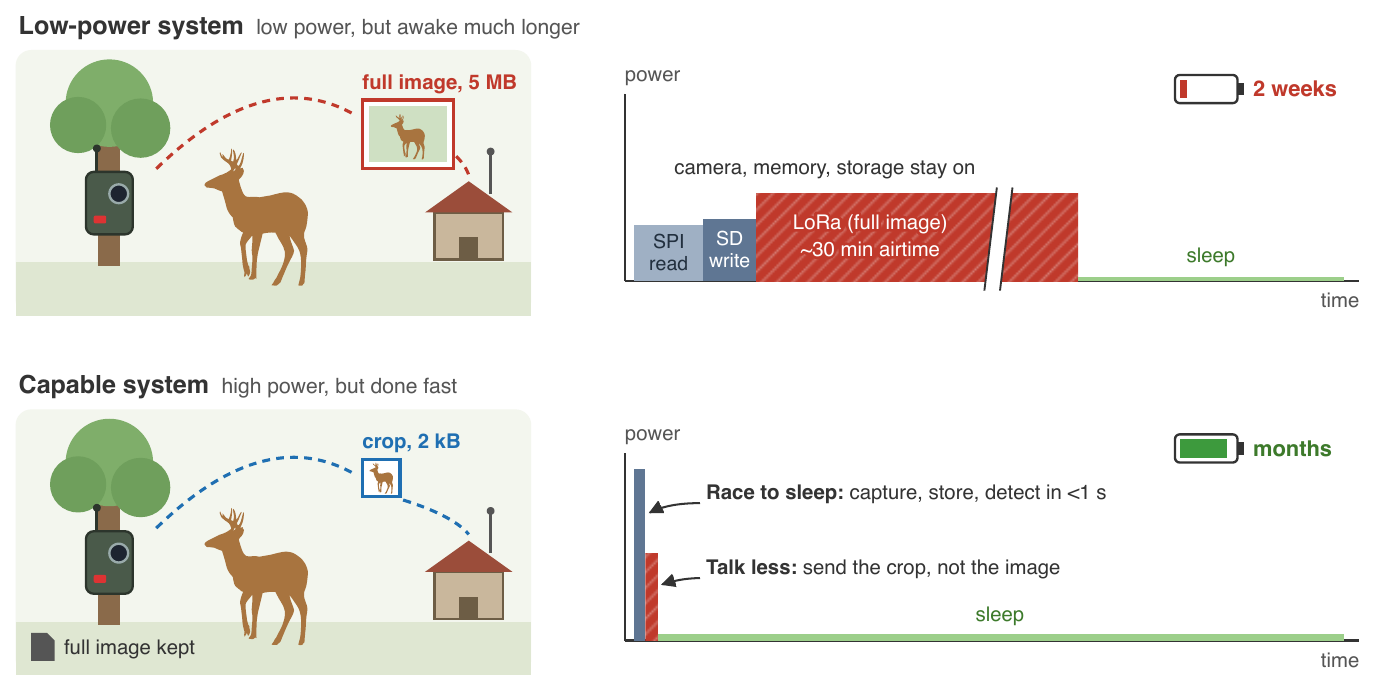}
  \caption{Low power is not low energy. A low-power system stays awake far longer and sends the full image. A capable one finishes fast and transmits less, so it uses less energy overall.}
  \Description{Two rows compare a camera trap built on a low-power microcontroller with one built on a capable microcontroller. Each row shows a trap on a tree photographing a deer and sending data to a ranger station, next to a plot of power over time. In the top row, the trap sends a 5 MB full image; its plot shows low but long SPI read and SD write blocks, then a very long LoRa transmission of about half an hour, then sleep, and a nearly empty battery labeled two weeks. In the bottom row, the trap sends a 2 kB crop and keeps the full image on its SD card; its plot shows one tall, narrow active block labeled race to sleep, a narrow LoRa block labeled talk less, then a long sleep, and a full battery labeled months.}
  \label{fig:teaser}
\end{teaserfigure}

\maketitle

\input{1-Introduction}
\input{3-Design}
\input{5-Evaluation}
\input{2-Related-Work}
\input{6-Conclusion}

\bibliographystyle{ACM-Reference-Format}
\bibliography{references}

\end{document}

%% file: globals.tex
\newcommand{\prh}[1]{\noindent\textbf{#1:}}

%% file: 0-Abstract.tex
\begin{abstract}
For two decades, long-lived sensor nodes have been built from the lowest-power parts, because their scalar readings take only milliseconds to acquire and send. Scientists now want higher fidelity data from the field, such as images and audio, and these observations break that rule: a low-power microcontroller stays awake for seconds to read, store, and transmit each one, so low power is no longer low energy. We argue that long-lived, high-fidelity nodes should instead be built around a more capable microcontroller, provided the whole board sleeps deeply. We revisit two principles from early sensor networks, \emph{race to sleep} and \emph{talk less}, for high-fidelity data and on-device AI, and test them by comparing the STM32N6, which pairs a fast core with a neural accelerator, against the low-power Apollo4 Plus. Although the STM32N6 draws 1.9$\times$ more power, it completes an observation (capture, storage, on-board detection, and a LoRa transmission) 9.9$\times$ faster and so uses 5.2$\times$ less energy. On the same model, it runs inference with 3.2--11.4$\times$ less energy, and it fits models over 23 points more accurate. A high-fidelity node lives longest when it finishes quickly, sleeps deeply, and sends only what matters.
\end{abstract}

%% file: 1-Introduction.tex
\section{Introduction}
\label{sec:intro}

% Alice is a park ranger who wants to know which animals live in his park throughout the year. She builds a camera trap around an ultra-low-power microcontroller, packs it with batteries, and programs it to send each image over LoRa, a low-power radio, to a ranger station more than a mile away. For two weeks, crisp photos arrive every day. Then the trap goes silent, and its batteries are dead. Alice is puzzled: \textbf{how did a node built from a low-power microcontroller and a low-power radio run out of energy so quickly?}

Alice is a park ranger. For years, her temperature and humidity sensors have run on a low-power microcontroller and a LoRa radio. To learn which animals live in the park, she adds a camera to the same platform and has it send each photo to a ranger station more than a mile away. For two weeks, crisp photos arrive every day. Then the platform goes silent, its batteries dead. Alice is puzzled: \textbf{how did a platform that lasted for years on temperature readings run out of energy in two weeks?}

Alice followed a design rule that has served sensor networks for two decades. Platforms such as Mica~\cite{hill2002mica} and Telos~\cite{polastre2005telos} made long-lived field sensing practical, and later nodes ran entirely on harvested RF power~\cite{sample2008wisp,menon2022wireless}. These nodes collected scalar data, such as temperature and humidity. A reading fits in a few bytes, so a node can acquire, store, and send it within milliseconds and spend nearly all of its life asleep. When the active time is this short, choosing the lowest-power parts is the same as choosing the lowest-energy ones.

Scientists now want higher-fidelity observations from the field. Images and audio, for example, support ecosystem monitoring and help detect events such as wildfires, and identify faults on remote infrastructure~\cite{beery2018recognition,desai2022camaroptera,hill2018audiomoth}. These observations are orders of magnitude larger than a scalar reading. Moving and preserving them takes more bandwidth and storage, and interpreting them in the field takes far more computation.

\textbf{Low power is not low energy.} Energy is power integrated over time, and a high-fidelity observation keeps the node awake far longer than a scalar reading. Alice's microcontroller draws little power, but it reads each image from the camera over a slow bus and writes it to storage in pieces, while the camera, memory, and storage stay powered (Figure~\ref{fig:teaser}). The radio is worse: at LoRa's data rate, a single 5~MB image takes about half an hour of airtime (\S\ref{sec:talk}). In the end, every part was low-power, but each observation became high-energy.

A more powerful microcontroller draws more current, but if it finishes each observation many times sooner, it can use less energy overall. With a neural accelerator, it could also interpret the image on board and send only what end-users need, such as a class label or a cropped segment, instead of the full sample. The catch is that racing pays off only if the node sleeps deeply. Application processors such as the Raspberry Pi are fast but idle at hundreds of milliwatts. New microcontrollers such as the STM32N6~\cite{st_stm32n6} and the MAX78000~\cite{adi_max78000} may offer both, pairing computing power and speed previously unheard of in microcontroller-class hardware with a microamp standby mode.

This paper takes the position that the platform for long-lived, high-fidelity sensing should be built around a more capable microcontroller, not the lowest-power one, provided the whole board sleeps deeply. We ground this position in two ideas from early sensor networks, \emph{race to sleep} and \emph{talk less}, revisited for high-fidelity data and on-device AI (\S\ref{sec:design}). To test them, we take the STM32N6 as a representative capable microcontroller and compare it with a low-power one, the Ambiq Apollo4 Plus (\S\ref{sec:eval}). Although the STM32N6 draws 1.9$\times$ more average power than the Apollo4 Plus while active, it finishes a full observation (capture, storage, on-board detection, and a LoRa transmission) 9.9$\times$ faster and so uses 5.2$\times$ less energy. Similarly, when running the same AI model, the STM32N6 draws 35 to 48$\times$ more power but runs 132 to 545$\times$ faster, so it uses 3.2 to 11.4$\times$ less energy. It also fits much larger models, which are over 23 and 24 points more accurate on ImageNet and COCO datasets, respectively.

These results support our position: for high-fidelity sensing, a node lives longest not when it is built from the lowest-power parts, but when it finishes its work quickly, sleeps deeply, and sends only what matters. Whether it listens for birdsong, watches for wildfire smoke, or inspects an aircraft for faults, such a node could run for months instead of weeks, deliver what scientists need from each observation, and keep every full observation archived on board.

%% file: 3-Design.tex
\section{Design Principles}
\label{sec:design}

We adapt two design principles for long-lived, high-fidelity sensing nodes. Both revisit ideas from early sensor networks~\cite{polastre2005telos,pottie2000wins} for high-fidelity data.

\subsection{Race to Sleep}
\label{sec:race}

A node should finish each observation quickly and then turn everything off. Consider a deployment that lasts $T$ seconds and records $N$ observations. Each event keeps the node active for $t_{\mathrm{active}}$ seconds, from wake-up until it returns to standby. The total energy, in joules, is
\begin{align}
E_{\mathrm{total}} &= P_{\mathrm{standby}}\,(T-N\,t_{\mathrm{active}}) + N\,E_{\mathrm{active}}, \label{eq:energy}\\
E_{\mathrm{active}} &= (P_{\mathrm{mcu}} + P_{\mathrm{rest}})\,t_{\mathrm{active}}, \nonumber
\end{align}
where $E_{\mathrm{active}}$ is the energy of one event. $P_{\mathrm{standby}}$ is the power of the whole board in standby, $P_{\mathrm{mcu}}$ is the power of the processor while active, and $P_{\mathrm{rest}}$ is the power of the rest of the board while active, such as the sensor, memory, storage, and regulators. The model assumes that every event takes the same time and energy. All powers are averages in watts.

For scalar data, $t_{\mathrm{active}}$ is a few milliseconds on any processor. For high-fidelity data, it is set by how fast the node moves, holds, and interprets the data, and the gap between processors is large. An uncompressed 5~MP image in RAW8, 5.04~MB, takes 4.0~s on a 10~Mbit/s SPI bus but about 40~ms on a 1~Gbit/s link. Moreover, this observation might be larger than the processor's SRAM, so it must be streamed in chunks, adding overhead, to external storage. Finally, if the node runs an AI model, the inference time can be hundreds of milliseconds to seconds on a low-power microcontroller but only a few milliseconds on a neural accelerator~\cite{lin2020mcunet}.

Throughout this $t_{\mathrm{active}}$, the node pays $P_{\mathrm{rest}}$ as well as $P_{\mathrm{mcu}}$, and a high-fidelity sensor such as a megapixel camera can draw tens to hundreds of milliamps while it streams. A node that draws four times the power but finishes ten times sooner therefore uses 0.4 times the active energy.

Racing pays off only if the node then sleeps deeply. When observations are rare, the standby term of Equation~\ref{eq:energy} dominates. A capable node also carries more parts, such as external memory and a high-speed sensor, each of which can leak in standby. Low standby power must therefore hold for the whole board, not only for the processor.

\subsection{Talk Less}
\label{sec:talk}

For scalar sensing, sending data is usually cheap. A reading takes milliseconds to send over a slow link such as LoRa and costs little over a metered cellular or satellite link. Raw high-fidelity data is different, because every observation is large. Over LoRa, a single 5~MP RAW8 image takes about half an hour of airtime (Table~\ref{tab:comm}). As a result, some of the most successful high-fidelity field devices, such as AudioMoth~\cite{hill2018audiomoth} and most commercial camera traps, do not transmit their data at all. They store it until someone visits the site.

On-device computation allows sending only what matters. A camera trap can send a crop of the animal instead of the whole scene, and an acoustic node can send a bird's call instead of minutes of recording. Table~\ref{tab:comm} shows that even with one second of on-device processing, sending a small result instead of the raw data cuts transmission time by up to $1{,}061\times$.

Trading computation for communication is a well-established idea in sensor networks~\cite{pottie2000wins}, and semantic communication takes it further by sending only task-relevant content~\cite{gunduz2023beyond,hojjat2024limitnet,huang2022real}. In these systems, the more the node can compute on board, the more accurately it can decide what matters and the less it must send. A fast accelerator also keeps the cost of that computation small, so racing to sleep and talking less reinforce each other.

\begin{table}[t]
\caption{Time to send a 5.04~MB raw observation (one 5~MP RAW8 frame) versus one second of processing plus a 2~kB result, at nominal rates~\cite{semtech_lr2021,iridium_9770,nordic_lte_rates}.}
\label{tab:comm}
\centering
\small
\begin{tabular}{@{}lrrrr@{}}
\toprule
Link   & Rate (kbps)   & Raw (s) & Result (s) & Speedup \\
\midrule
LoRa SF7   & 22  & 1832.3 & 1.73 & $1061\times$ \\
Iridium   & 22  & 1832.3 & 1.73 & $1061\times$ \\
NB-IoT    & 60   & 671.8  & 1.27 & $530\times$  \\
LTE-M     & 375  & 107.5  & 1.04 & $103\times$  \\
\bottomrule
\end{tabular}
\end{table}

Talking less does not mean discarding data. The node can archive every original locally and send only what is useful now.

%% file: 5-Evaluation.tex
\section{Evaluation}
\label{sec:eval}

We test the two principles on two microcontrollers that represent the choice a node designer faces: an ultra-low-power one and a more capable one with a neural accelerator. We ask what it costs each to acquire and store a high-fidelity observation, how that cost translates into battery life, and which models each can run on that observation.

\subsection{Setup}
\label{sec:systems}

\begin{table}[b]
  \caption{Systems under test. Values are from datasheets.}
  \label{tab:systems}
  \centering
  \small
  \setlength{\tabcolsep}{3pt}
  \begin{tabular}{@{}lcc@{}}
    \toprule
    & Apollo4 Plus & STM32N6 \\
    \midrule
    Core & M4F, 192~MHz & M55, 800~MHz \\
    Accelerator & None & Neural-ART NPU \\
    Memory & 2.75~MB SRAM & 4.2~MB SRAM, 32~MB PSRAM \\
    Storage & 2~MB MRAM & 64~MB NOR Flash\\
    Camera link & SPI & MIPI CSI-2 \\
    Sleep power & 50.9~$\mu$W & 262.62~$\mu$W \\
    \bottomrule
  \end{tabular}
\end{table}

Table~\ref{tab:systems} lists the two systems. The \textbf{Apollo4 Plus}~\cite{ambiq_apollo4} is an ultra-low-power microcontroller with a 192~MHz Cortex-M4F, which we use on Ambiq's evaluation board (AMAP4PEVB). It has no camera interface, so it reads a 5~MP SPI Arducam camera module with OV5642 image sensor. The \textbf{STM32N6}~\cite{st_stm32n6} (STM32N657) pairs an 800~MHz Cortex-M55 with a neural accelerator. We use it as a representative capable microcontroller on a board we built that receives frames over MIPI CSI-2 by DMA into external PSRAM. It pairs with the Arducam Sony IMX335 image sensor. Both systems use a Semtech LR2021 LoRa radio at 915~MHz with spreading factor 7, 500~kHz bandwidth and +14~dBm transmit power.

Both systems capture a 2592$\times$1944 image and store it on a microSD card in RAW8 format. Each runs at its fastest clock, uses DMA where available, and enters its lowest-power state that retains enough state to resume. We report board-level power for both systems, measured with a Joulescope JS220 at 1000~kHz and averaged over 10 runs, while connected to 4 AA alkaline cells. For the STM32N6, this is the whole board measured at the battery, including its regulators, external memory, and every other part. The Ambiq evaluation board carries a debugger, indicator LEDs, and other parts that a field node would not have, so for the Apollo4 Plus we measure major components separately---the processor, image sensor, microSD card, and LoRa radio---and report their sum. This sum omits the regulator losses and leakage that a complete Apollo4 Plus board would add, so the comparison favors the Apollo4 Plus.

\subsection{End-to-End Observation}
\label{sec:eval-capture}

Each system starts in its low-power state, wakes, captures one 5~MP image, writes it to microSD, runs an object detection model on it, transmits the crop image over LoRa radio, and returns to its low-power state (Table~\ref{tab:capture}).

\begin{table}[t]
  \caption{Time and board-level energy while waking, processing, and transmitting one 5~MP image. Some stages overlap, so total values are not exact sums of individual stages.}
  \label{tab:capture}
  \centering
  \small
  \setlength{\tabcolsep}{8pt}
  \begin{tabular}{@{}lrrrr@{}}
    \toprule
    & \multicolumn{2}{c}{Apollo4 Plus} & \multicolumn{2}{c}{STM32N6} \\
    \cmidrule(lr){2-3}\cmidrule(l){4-5}
    Stage & ms & mJ & ms & mJ \\
    \midrule
    Wake & 2.1 & 0.4 & 4.7 & 1.7 \\
    Capture to memory & 1012.3 & 520.6 & 214.3 & 129.6 \\
    Write to microSD & 5142.4 & 4040.0 & 291.8 & 72.5 \\
    ML inference & 17139.5 & 638.8 & 446.1 & 159.7 \\
    LoRa TX (ROI) & 1855.9 & 291.7 & 1844.0 & 692.3 \\
    %Re-arm standby & 44.3 & 25.8 & 28.2 & 10.5 \\
    \midrule
    \textbf{Active total} & 25196.5 & 5517.4 & 2537.2 & 1066.2 \\
    Scale (time)   & 9.9$\times$ & & 1$\times$ & \\
    Scale (energy) & & 5.2$\times$ & & 1$\times$ \\
    \bottomrule
  \end{tabular}
\end{table}

\prh{Active time and energy} The Apollo4 Plus needs 1,012.3~ms to read a 5.04~MB frame over SPI, while the camera and processor stay on. Over MIPI CSI-2, the STM32N6 moves the frame into memory in 214.3~ms. Over the whole event, the STM32N6 draws 1.9$\times$ more average power than the Apollo4 Plus while active (421 vs.\ 219~mW) but finishes 9.9$\times$ sooner, so it uses 5.2$\times$ less energy per image.

% \prh{Standby} With the image sensor, microSD card, and radio switched off, the Apollo4 Plus draws [X]~$\mu$A. The STM32N6 board draws 1241~$\mu$A, although the STM32N6 itself is specified at [X]~$\mu$A in Standby~\cite{st_stm32n6}. The rest is drawn by \todo{what leaks: regulators, PSRAM, pull-ups?}. This gap shows how easily the parts around a capable processor can erase its deep sleep.

\subsection{Lifetime Projection}
\label{sec:eval-life}

Setting $T = 86{,}400$~s (one day) and $N = n$ captures per day in Equation~\ref{eq:energy} gives the daily energy, in joules,
\begin{equation}
E_{\mathrm{day}} = P_{\mathrm{standby}}\,(86{,}400~\mathrm{s} - n\,t_{\mathrm{active}}) + n\,E_{\mathrm{active}},
\label{eq:daily}
\end{equation}
where $t_{\mathrm{active}}$ (s) and $E_{\mathrm{active}}$ (J) are the measured active time and energy per capture (Table~\ref{tab:capture}), %and $P_{\mathrm{standby}}$ (W) is the measured standby current times the [X]~V supply voltage.
and $P_{\mathrm{standby}}$ (W) is the standby power of the microcontroller from datasheets in Table~\ref{tab:systems}. A well-designed board would draw only slightly more than its microcontroller, so the datasheet value gives an optimistic but representative trend. Lifetime, in days, is $E_{\mathrm{batt}}/E_{\mathrm{day}}$, where $E_{\mathrm{batt}} = 49{,}250$~J (13.7~Wh) is the energy of four Duracell alkaline AA cells (2.85~Ah at an average 1.2~V each~\cite{duracell_mn1500}).

Table~\ref{tab:daily} applies it to 1, 10, 50, and 100 captures per day. At one capture a day, standby is 96\% of the STM32N6's daily energy, so it uses 2.4$\times$ more energy than the Apollo4 Plus. As the rate grows, active energy dominates, and at 100 captures a day the STM32N6 uses 4.3$\times$ less. Equating $E_{\mathrm{day}}$ for the two microcontrollers, and neglecting standby energy during the short active periods ($n\,t_{\mathrm{active}} \ll 86{,}400$~s), gives the break-even capture rate
\begin{equation}
n^{*} \approx \frac{(P^{\mathrm{N6}}_{\mathrm{standby}} - P^{\mathrm{A4}}_{\mathrm{standby}}) \cdot 86{,}400~\mathrm{s}}{E^{\mathrm{A4}}_{\mathrm{active}} - E^{\mathrm{N6}}_{\mathrm{active}}}
\label{eq:crossover}
\end{equation}
in captures per day, where superscripts N6 and A4 denote the STM32N6 and the Apollo4 Plus. With datasheet standby values, $n^{*}$ is 4 captures a day. Above that, the STM32N6 has longer battery life than the Apollo4 Plus.

\begin{table}[t]
  \caption{Daily energy (J) and lifetime (days, in parentheses) on four AA alkaline cells, projected using measured active energy and datasheet standby power.}
  \label{tab:daily}
  \centering
  \small
  \setlength{\tabcolsep}{3.5pt}
  \begin{tabular}{@{}lrrrr@{}}
    \toprule
    Captures/day & 1 & 10 & 50 & 100 \\
    \midrule
    Apollo4 Plus & 9.9 (4968) & 59.6 (827) & 280.2 (176) & 556.0 (89) \\
    STM32N6 & 23.8 (2073) & 33.4 (1476) & 76.1 (648) & 129.4 (381) \\
    \bottomrule
  \end{tabular}
\end{table}

\begin{figure*}[t]
  \centering
  \includegraphics[width=\textwidth]{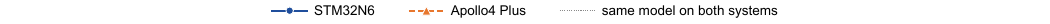}
  \begin{minipage}[b]{0.25\textwidth}
    \centering
    \includegraphics[width=\linewidth]{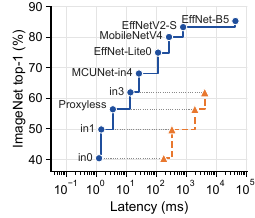}\\
    {\small (a) ImageNet, latency}
  \end{minipage}\hfill
  \begin{minipage}[b]{0.25\textwidth}
    \centering
    \includegraphics[width=\linewidth]{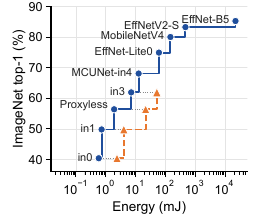}\\
    {\small (b) ImageNet, energy}
  \end{minipage}\hfill
  \begin{minipage}[b]{0.25\textwidth}
    \centering
    \includegraphics[width=\linewidth]{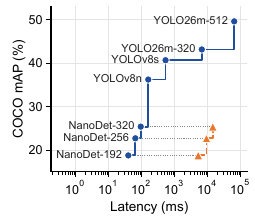}\\
    {\small (c) COCO, latency}
  \end{minipage}\hfill
  \begin{minipage}[b]{0.25\textwidth}
    \centering
    \includegraphics[width=\linewidth]{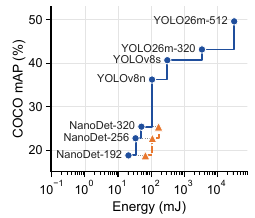}\\
    {\small (d) COCO, energy}
  \end{minipage}
   \caption{Accuracy versus measured latency and energy per inference, with each system's Pareto front. STM32N6 energy is whole-board, while Apollo4 Plus energy is MCU-only.}
  \Description{Four scatter plots in a row, each with a logarithmic horizontal axis and accuracy on the vertical axis. Panels (a) and (b) show ImageNet top-1 accuracy against latency and against energy per inference. Panels (c) and (d) show COCO mAP against latency and against energy. In each panel, the STM32N6 and the Apollo4 Plus each form a staircase-shaped Pareto front. The Apollo4 Plus runs only the smallest models: four classifiers up to 62\% top-1 and three NanoDet detectors up to 25\% mAP. The STM32N6 runs models up to 85\% top-1 and 50\% mAP. Dotted horizontal lines join each model that runs on both systems. The STM32N6 is 132 to 545 times faster on the same model and uses 3.2 to 11.4 times less energy, even though its energy includes the whole board.}
  \label{fig:ml}
\end{figure*}
\subsection{On-Device Inference}
\label{sec:eval-ml}

To measure how much model capacity each system offers, we use standard models on standard benchmarks: image classification on ImageNet (top-1 on all 50,000 validation images)~\cite{deng2009imagenet} and object detection on COCO (mAP over IoU thresholds from 0.5 to 0.95 on the 5,000 validation images)~\cite{lin2014microsoft}. All models are quantized to INT8, and we measure accuracy on a host for the exact file each system runs. The models run on the NPU of the STM32N6 with ST Edge AI, and on the CPU of the Apollo4 Plus with TensorFlow Lite Micro and CMSIS-NN. For each system, we compile every model that fits its memory, measure its latency and energy per inference on the device (averaged over 10 inferences), and plot the Pareto front of accuracy against latency and against energy (Figure~\ref{fig:ml}). STM32N6 energy is for the whole board at the battery, while Apollo4 Plus energy is for its MCU supply only, which favors the Apollo4 Plus.

\prh{Bigger models fit} The Apollo4 Plus is limited by its 2.75~MB of SRAM for activations and its 2~MB of MRAM for weights. For classification, its most accurate model is MCUNet-in3 at 62.0\% top-1, while the STM32N6 runs EfficientNet-B5 (Noisy Student) at 85.3\%, 23.3 points higher (Figure~\ref{fig:ml}a). The STM32N6 holds the larger weights in external PSRAM and flash. For detection, the Apollo4 Plus front ends at NanoDet-Plus-m at 320$\times$320, which reaches 25.4 mAP, while the STM32N6 front extends to YOLO26m at 512$\times$512, which reaches 49.6 mAP, 24.2 points higher (Figure~\ref{fig:ml}c).

\prh{The same model costs less} Every model that fits the Apollo4 Plus also runs on the STM32N6, so comparing them isolates the hardware (dotted lines in Figure~\ref{fig:ml}). For classification, MCUNet-in3 takes 4.15~s on the Apollo4 Plus and 13.8~ms on the STM32N6, 300$\times$ faster. The STM32N6 draws 41$\times$ more power while active (515 vs.\ 12.5~mW) but uses 7.3$\times$ less energy per inference (7.1 vs.\ 51.7~mJ), even though its energy includes the whole board; across the four classifiers, the gap is 4.1 to 11.4$\times$ (Figure~\ref{fig:ml}b). For detection, NanoDet-Plus-m at 320$\times$320 takes 14.4~s on the Apollo4 Plus and 94~ms on the STM32N6, 153$\times$ faster, and uses 3.3$\times$ less energy (51 vs.\ 169~mJ; Figure~\ref{fig:ml}d), with a similar 3.2 to 3.3$\times$ gap for the smaller NanoDet-Plus-m inputs. Across the seven models that fit both systems, the STM32N6 is 132 to 545$\times$ faster at the same accuracy.
% \subsection{Reporting Results}
% \label{sec:eval-report}

% Finally, we measure the energy of reporting one observation over LoRa (Table~\ref{tab:report}). A node can send the full JPEG image, the crop of each animal found by the detector, or only the label from the classifier. We transmit each payload at SF7 \todo{radio part, TX power, supply voltage}, measure the board-level energy of each transmission, and add the measured energy of capturing, storing, and interpreting the image from Table~\ref{tab:capture} and Figure~\ref{fig:ml}.

% \begin{table}[t]
%   \caption{Energy to report one observation over LoRa SF7, including the measured energy of capture, storage, and inference.}
%   \label{tab:report}
%   \centering
%   \small
%   \setlength{\tabcolsep}{3pt}
%   \begin{tabular}{@{}lrrrrr@{}}
%     \toprule
%     & & & & \multicolumn{2}{c}{Total (J)} \\
%     \cmidrule(l){5-6}
%     Payload & Size & Airtime & Radio (J) & A4P & N6 \\
%     \midrule
%     Full image & [X]~kB & [X]~s & [X] & [X] & [X] \\
%     Detection crop & [X]~kB & [X]~s & [X] & [X] & [X] \\
%     Class label & [X]~B & [X]~s & [X] & [X] & [X] \\
%     \bottomrule
%   \end{tabular}
% \end{table}

% Sending the full image costs [X]~J of radio energy and [X]~minutes of airtime \todo{note regional duty-cycle or dwell-time limits}. Sending a crop instead reduces the total by [X]$\times$ on the STM32N6. On the Apollo4 Plus, the slower inference consumes [X]\% of the savings.

%% file: 2-Related-Work.tex
\section{Related Work}
\label{sec:related-work}

\paragraph{Low-power sensor nodes.} Motes such as Telos~\cite{polastre2005telos} and ENTS~\cite{madden2026ents} run for months to years, but they were built for scalar sensors. Open acoustic recorders such as AudioMoth~\cite{hill2018audiomoth} show how much an affordable, open, long-lived device can do for ecology, but they do little analysis on board. Commercial camera traps last for months on AA batteries, but they are closed devices that only record~\cite{reconyx_hyperfire2,bushnell_core_ds4k}. TinyML runs small networks on microcontrollers~\cite{lin2020mcunet}, but limited memory restricts it to small models on low-resolution inputs~\cite{lee2025hypercam}.

\paragraph{Low-power imaging.} Imaging systems built on low-power microcontrollers fit the workload to the processor. WISPCam captures low-resolution frames on harvested RF power and sends them to a nearby reader~\cite{naderiparizi2015wispcam}. Camaroptera harvests energy, runs a small network on low-resolution images, and sends only the interesting ones over LoRa~\cite{desai2022camaroptera}. SONIC and TAILS make such inference correct and efficient under intermittent power~\cite{gobieski2019intelligence}. These designs reach long lifetimes, but part of that lifetime comes from what the node does not capture or analyze.

\paragraph{Embedded vision hardware.} Field camera traps built on single-board computers such as the Raspberry Pi and Jetson Nano run full vision models and send only labels, over LoRaWAN or satellite~\cite{zualkernan2022iot,whytock2023realtime}. Their power draw of hundreds of milliwatts or more requires solar panels or large batteries. Accelerator microcontrollers such as the MAX78000 run small networks quickly, but they hold under half a megabyte of weights~\cite{adi_max78000}. Ultra-low-power parts such as the Apollo4 Plus hold megabytes of SRAM at microamp sleep currents, but they run models on a general-purpose core~\cite{ambiq_apollo4}. The STM32N6 combines a neural accelerator, megabytes of SRAM, a MIPI camera interface, and a microamp-level standby mode~\cite{st_stm32n6}. Parts like it make a capable, long-lived node feasible, but the board around the chip determines whether that node actually sleeps deeply.

%% file: 6-Conclusion.tex
\section{Conclusion}

Long-lived sensing was built on the rule of choosing the lowest-power parts. That rule held for scalar readings, which kept a node awake for only milliseconds, but it no longer holds for higher-fidelity data such as images, audio, and video. Comparing a capable microcontroller, the STM32N6, with a low-power one, the Apollo4 Plus, we found that racing to sleep pays off. The STM32N6 draws more power but finishes a full observation 9.9$\times$ faster and uses 5.2$\times$ less energy, and it outlasts the Apollo4 Plus above about four captures a day. Its accelerator also makes talking less practical. It runs larger and more accurate models with less energy, so a node can send a crop of what it saw instead of the image itself. These gains hold only if the whole board sleeps as deeply as its processor. We are building our board as an open platform toward that goal, to unlock field research at scales and fidelities that are out of reach today.